\documentclass[12pt]{article}
\usepackage{amsmath,amsfonts,bbold,hyperref,graphicx}

\newcommand{\be}{\begin{equation}}
\newcommand{\ee}{\end{equation}}
\newcommand{\bea}{\begin{eqnarray}}
\newcommand{\eea}{\end{eqnarray}}
\newcommand{\beas}{\begin{eqnarray*}}
\newcommand{\eeas}{\end{eqnarray*}}

\def\identity{{\mathbb{1}}}
\def\slash#1{\setbox0=\hbox{#1}\copy0\kern-\wd0\hbox to \wd0{\hss \, \big/ \hss}}

\begin{document}
\begin{titlepage}

\begin{center}

{\Large Anomalies and Bose symmetry}

\vspace{12mm}

\renewcommand\thefootnote{\mbox{$\fnsymbol{footnote}$}}

Daniel Kabat

\vspace{2mm}

{\small \sl Department of Physics and Astronomy} \\
{\small \sl Lehman College, City University of New York, Bronx NY 10468, USA}

\vspace{1mm}

{\small \tt daniel.kabat@lehman.cuny.edu}

\end{center}

\vspace{8mm}

\noindent
We point out a feature of the triangle diagram for three chiral currents which is perhaps not widely appreciated: Bose symmetry is not manifest and suffers from a momentum-routing ambiguity.
Imposing Bose symmetry fixes the ambiguity and leads to the famous Adler - Bell - Jackiw anomaly.

\vspace{5cm}
\centerline{\em Dedicated to Roman Jackiw on the occasion of his 80${}^{\,th}$ birthday}

\end{titlepage}
\setcounter{footnote}{0}
\renewcommand\thefootnote{\mbox{\arabic{footnote}}}

\section{Introduction}

It was a privilege studying under Roman Jackiw at MIT in the early 90's.  By that time the threat of midnight phone calls inquiring after the status of a
calculation had somewhat subsided, but the benefits of Roman as an advisor remained.  Never one to waste time with trivial discussion or unfounded speculation, he instead provided
an unparalleled source of clear guidance and direction for his students.

Roman had an uncanny ability to formulate and solve mathematical systems of physical relevance.  His work often laid the foundation for future developments, with an impact far
beyond its original scope.
Witness the resurgence of interest in Jackiw - Teitelboim gravity \cite{Jackiw:1984je,Teitelboim:1983ux} as a holographic system \cite{Almheiri:2014cka,Maldacena:2016upp,Engelsoy:2016xyb}, or the Adler - Bell - Jackiw anomaly \cite{Adler:1969gk,Bell:1969ts} which
had its origins in current algebra but then did so much to initiate the use of topological methods in gauge theories \cite{Jackiw:1983nv}.  Another example may be his work on non-associative structures
and 3-cocycles \cite{Jackiw:1984rd}, a topic I believe will ultimately play an important role in quantizing gravity \cite{Kabat:2018pbj}.

In these notes I'll consider a two-component fermion of definite chirality.  Thus in section \ref{sect:chiral} I'll work with a single chiral current, instead of the vector and axial combinations which can be built from a pair of chiral fermions, and
I'll discuss a feature of the triangle diagram for three chiral currents which, although known, may not be widely appreciated: due to a momentum-routing ambiguity the diagram does not have manifest Bose symmetry.  Bose symmetry may be imposed on the diagram by hand; this fixes the ambiguity and leads to a unique expression for the divergence of the chiral current.  In section \ref{sect:axial} I'll make contact with the usual axial anomaly.  In section \ref{sect:counterterm} I'll show that Bose symmetry can be restored with a local counterterm, while a different counterterm gives a covariant expression for the anomaly.

\section{The chiral triangle\label{sect:chiral}}

Consider a massless chiral fermion coupled to an external vector field.  We'll describe the fermion using a Dirac spinor $\psi$, but with a projection condition so that the fermion is either right- or left-handed.\footnote{Conventions: the metric $g_{\mu\nu} = {\rm diag} (+---)$, the antisymmetric tensor $\epsilon_{0123} = +1$,
and the Dirac matrices are
\[
\gamma^0 = \left(\begin{array}{cc} 0 & \identity \\ \identity & 0 \end{array}\right) \qquad\quad
\gamma^i = \left(\begin{array}{cc} 0 & \sigma^i \\ -\sigma^i & 0 \end{array}\right) \qquad\quad
\gamma^5 \equiv i \gamma^0 \gamma^1 \gamma^2 \gamma^3 = \left(\begin{array}{cc} - \identity & 0 \\ 0 & \identity \end{array}\right)
\]
In this basis a Dirac spinor has left- and right-handed chiral components $\psi = \left({\psi_L \atop \psi_R}\right)$.}
That is, we'll assume
\be
\psi = {1 \over 2} \big(1 \pm \gamma^5\big) \psi
\ee
{\em In what follows the upper sign will correspond to a right-handed spinor, the lower sign to left-handed.}  The coupling of $\psi$ to the external (non-dynamical) vector field $A_\mu$ is described by the Lagrangian
\be
\label{mcse}
{\cal L} = \bar{\psi} i \gamma^\mu \left(\partial_\mu + i q A_\mu\right)\psi
\ee

With these ingredients the diagram for scattering three vector fields
\begin{center}
\raisebox{-2.25cm}{\includegraphics[height=4.8cm]{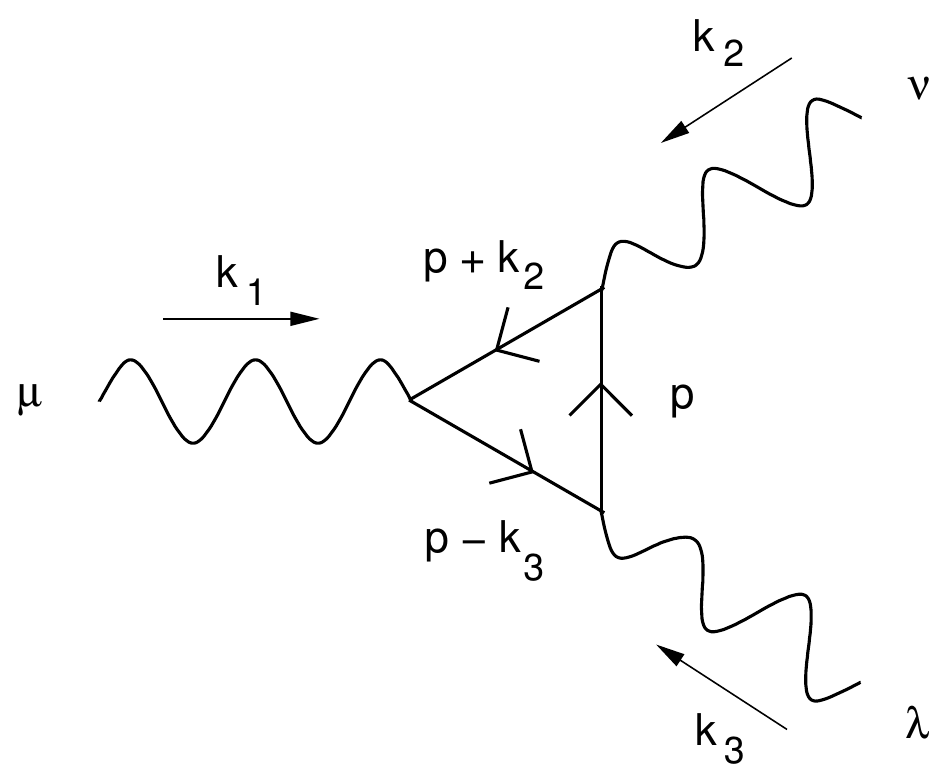}} + \qquad
\parbox{4cm}{crossed diagram \\[3pt] $(k_2,\nu) \leftrightarrow (k_3,\lambda)$}
\end{center}
leads to the scattering amplitude
\bea
\nonumber
-i{\cal M}_{\mu\nu\lambda} & = & (-1) \int {d^4p \over (2\pi)^4} {\rm Tr} \,
\left\lbrace -iq\gamma_\mu {i \over \slash{$p$} + \slash{$k$}_2}
\big(-iq\gamma_\nu\big) {i \over \slash{$p$}}
\big(-iq\gamma_\lambda\big) {i \over \slash{$p$} - \slash{$k$}_3}
\, {1 \over 2} \big(1 \pm \gamma^5\big)\right\rbrace \\
\label{TriangleAmplitude}
& & \qquad\quad + {\rm Tr} \,
\left\lbrace -iq\gamma_\mu {i \over \slash{$p$} + \slash{$k$}_3}
\big(-iq\gamma_\lambda\big) {i \over \slash{$p$}}
\big(-iq\gamma_\nu\big) {i \over \slash{$p$} - \slash{$k$}_2}
\, {1 \over 2} \big(1 \pm \gamma^5\big)\right\rbrace
\eea
All external momenta are directed inward, with $k_1 + k_2 +
k_3 = 0$.  The projection operator ${1 \over 2} \big(1 \pm \gamma^5\big)$ ensures that only a single chirality circulates in the loop.

We might expect the (necessarily chiral) currents $j^\mu = \bar{\psi} \gamma^\mu \psi$ to obey Bose statistics,
so we might expect the amplitude to be invariant under permutations of the external lines.
Indeed there's a simple argument which seems to show that Bose symmetry is
satisfied.  Invariance under exchange $(k_2,\nu) \leftrightarrow (k_3,
\lambda)$ is manifest; given our labellings it just corresponds to
exchanging the two diagrams.  However we should check invariance under
exchange of say $(k_1,\mu)$ with $(k_2,\nu)$.  Making this
exchange in (\ref{TriangleAmplitude}) we get
\beas
-i{\cal M}_{\nu\mu\lambda} & = & (-1) \int {d^4p \over (2\pi)^4} {\rm Tr} \,
\left\lbrace -iq\gamma_\nu {i \over \slash{$p$} + \slash{$k$}_1}
\big(-iq\gamma_\mu\big) {i \over \slash{$p$}}
\big(-iq\gamma_\lambda\big) {i \over \slash{$p$} - \slash{$k$}_3}
\, {1 \over 2} \big(1 \pm \gamma^5\big)\right\rbrace \\
& & \qquad\quad + {\rm Tr} \,
\left\lbrace -iq\gamma_\nu {i \over \slash{$p$} + \slash{$k$}_3}
\big(-iq\gamma_\lambda\big) {i \over \slash{$p$}}
\big(-iq\gamma_\mu\big) {i \over \slash{$p$} - \slash{$k$}_1}
\, {1 \over 2} \big(1 \pm \gamma^5\big)\right\rbrace
\eeas
Shifting the integration variable $p^\mu \rightarrow p^\mu + k_3^\mu$
in the first line, $p^\mu \rightarrow p^\mu - k_3^\mu$ in the
second, and making some cyclic permutations inside the trace, we seem
to recover the previous expression (\ref{TriangleAmplitude}).

Famously, though, this argument for Bose symmetry is invalid.  Instead a linearly divergent integral picks up a finite surface term when the integration variable is shifted.
\bea
\nonumber
\int {d^4p \over (2\pi)^4} \, f(p + a) & = & \int {d^4p \over (2\pi)^4} \, \big(f(p) + a^\mu \partial_\mu f(p) + \cdots \big) \\
\label{surface}
& = & \int {d^4p \over (2\pi)^4} \, f(p) -i a^\mu {1 \over 8\pi^2} \lim_{p \rightarrow \infty}
\langle p^2 p_\mu f(p) \rangle
\eea
Here angle brackets denote an average over the Lorentz group, and the limit is understood to mean large spacelike momentum.

We'll return below to see that the amplitude (\ref{TriangleAmplitude}) indeed violates Bose symmetry.
But for now, rather than study the violation of Bose
symmetry in detail, we're simply going to demand that the scattering
amplitude be symmetric.  The most straightforward way to do
this is to {\em define} the scattering amplitude to be given by
averaging over all permutations of the external lines.  Equivalently, we
average over cyclic permutations of the internal momentum routing.  That is,
we define the Bose-symmetrized amplitude
\beas
-i{\cal M}^{\rm symm}_{\mu\nu\lambda} & = & {1 \over 3} \, \Biggl[ \,
\raisebox{-1.2cm}{\includegraphics[height=2.4cm]{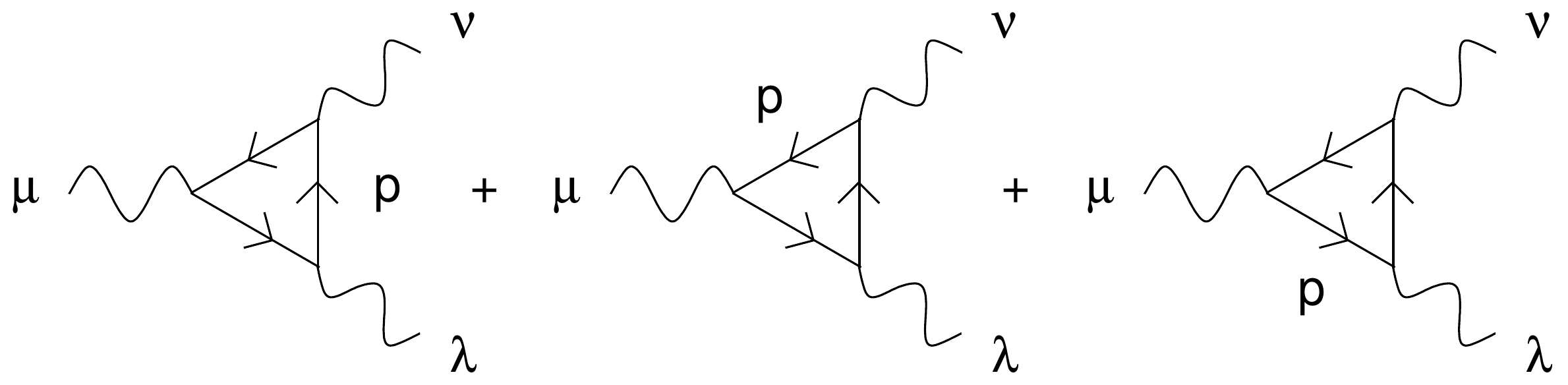}} \\
& & \quad\,\, + \,\,\, \parbox{3.4cm}{crossed diagrams \\[3pt] $(k_2,\nu) \leftrightarrow (k_3,\lambda)$} \, \Biggr]
\eeas
Explicitly this gives
\beas
-i{\cal M}^{\rm symm}_{\mu\nu\lambda} & = & \mp {1 \over 6} q^3
\int {d^4p \over (2\pi)^4} \, {\rm Tr} \, \biggl\lbrace
\gamma_\mu {1 \over \slash{$p$}} \gamma_\nu {1 \over \slash{$p$} -
\slash{$k$}_2} \gamma_\lambda {1 \over \slash{$p$} + \slash{$k$}_1} \gamma^5
+ \gamma_\mu {1 \over \slash{$p$} + \slash{$k$}_2} \gamma_\nu {1 \over
\slash{$p$}} \gamma_\lambda {1 \over \slash{$p$} - \slash{$k$}_3} \gamma^5
\\ & & \qquad\qquad
+ \gamma_\mu {1 \over \slash{$p$} - \slash{$k$}_1} \gamma_\nu {1 \over
\slash{$p$} + \slash{$k$}_3} \gamma_\lambda {1 \over \slash{$p$}} \gamma^5
+ \gamma_\mu {1 \over \slash{$p$}} \gamma_\lambda {1 \over \slash{$p$} -
\slash{$k$}_3} \gamma_\nu {1 \over \slash{$p$} + \slash{$k$}_1} \gamma^5
\\ & & \qquad\qquad
+ \gamma_\mu {1 \over \slash{$p$} + \slash{$k$}_3} \gamma_\lambda {1
\over \slash{$p$}} \gamma_\nu {1 \over \slash{$p$} - \slash{$k$}_2} \gamma^5
+ \gamma_\mu {1 \over \slash{$p$} - \slash{$k$}_1} \gamma_\lambda
{1 \over \slash{$p$} + \slash{$k$}_2} \gamma_\nu {1 \over \slash{$p$}}
\gamma^5 \biggr\rbrace
\eeas
Here we've used the fact that only terms involving $\gamma^5$
contribute to the scattering amplitude (Furry's theorem).

Having enforced Bose symmetry, let's check current conservation by
dotting this amplitude into $k_1^\mu$.  Using trivial identities such as
\be
\label{trivial}
\slash{$\displaystyle k$}_1 = \big(\slash{$\displaystyle p$} + \slash{$\displaystyle k$}_1\big) - \slash{$\displaystyle p$}
\ee
to cancel the propagators adjacent to $\slash{$k$}_1$, it
turns out that most terms cancel, leaving only
\beas
-i k_1^\mu {\cal M}^{\rm symm}_{\mu\nu\lambda} & = & \pm {1 \over 6}
q^3 \int {d^4p \over (2\pi)^4} \, {\rm Tr} \, \biggl\lbrace
{1 \over \slash{$p$} + \slash{$k$}_2} \gamma_\nu {1 \over \slash{$p$} -
\slash{$k$}_1} \gamma_\lambda \gamma^5
- {1 \over \slash{$p$} + \slash{$k$}_1} \gamma_\nu {1 \over \slash{$p$} -
\slash{$k$}_2} \gamma_\lambda \gamma^5 \\
& & \qquad\qquad
+ {1 \over \slash{$p$} - \slash{$k$}_1} \gamma_\nu {1 \over \slash{$p$} +
\slash{$k$}_3} \gamma_\lambda \gamma^5
- {1 \over \slash{$p$} - \slash{$k$}_3} \gamma_\nu {1 \over \slash{$p$} +
\slash{$k$}_1} \gamma_\lambda \gamma^5 \biggr\rbrace
\eeas
After shifting $p \rightarrow p + k_2 - k_1$ the second term seems to
cancel the first, and after shifting $p \rightarrow p + k_3 - k_1$ the
fourth term seems to cancel the third.  This naive cancellation means
the whole expression is given just by a surface term.
\beas
-i k_1^\mu {\cal M}^{\rm symm}_{\mu\nu\lambda} = \pm {1 \over 6}
q^3 \int {d^4p \over (2\pi)^4} \, (k_2^\alpha - k_1^\alpha) {\partial
\over \partial p^\alpha} {\rm Tr} \left\lbrace {1 \over \slash{$p$}}
\gamma_\nu {1 \over \slash{$p$} + \slash{$k$}_3} \gamma_\lambda \gamma^5
\right\rbrace &&\\
+ \left(k_3^\alpha - k_1^\alpha\right) {\partial \over
\partial p^\alpha} {\rm Tr} \left\lbrace {1 \over \slash{$p$} +
\slash{$k$}_2} \gamma_\nu {1 \over \slash{$p$}} \gamma_\lambda \gamma^5
\right\rbrace &&
\eeas
Using the expression for the surface term (\ref{surface}), evaluating the
Dirac traces with ${\rm Tr}\left(\gamma^\alpha\gamma^\beta\gamma^\gamma\gamma^\delta\gamma^5\right)
= 4 i \epsilon^{\alpha\beta\gamma\delta}$, and averaging over the Lorentz group with
$\langle p_\alpha p_\beta \rangle = {1 \over 4} g_{\alpha\beta} \, p^2$ we see that
the amplitude satisfies
\be
\label{anomaly}
-i k_1^\mu {\cal M}^{\rm symm}_{\mu\nu\lambda} = \mp {q^3 \over
12\pi^2} \epsilon_{\nu\lambda\alpha\beta} k_2^\alpha k_3^\beta\,.
\ee
Famously current conservation is violated by the triangle diagram \cite{Adler:1969gk,Bell:1969ts}.

This consistent anomaly \cite{Bardeen:1984pm} can be encapsulated by writing down an effective action for the vector field
$\Gamma[A]$ which incorporates the effect of the fermion triangle.  The
amplitude we've computed corresponds to the following non-local term in the effective action.\footnote{To verify (\ref{AnomalousEA})
note that the term we've written down in $\Gamma[A]$ corresponds to a vertex
\begin{center}
\hbox{\raisebox{-1.4cm}{\includegraphics[height=3.0cm]{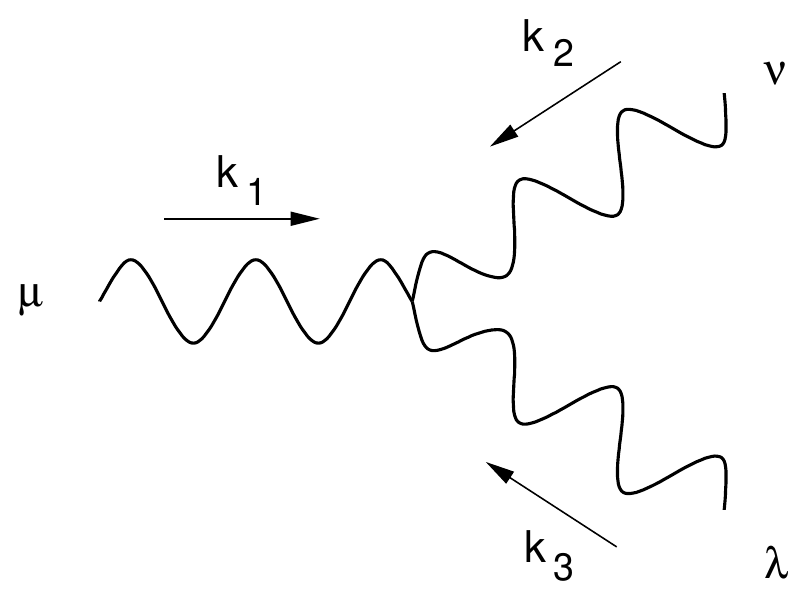}}
\qquad $\displaystyle \mp {q^3 \over 12\pi^2} \left({1 \over k_1^2} k_{1\mu} \epsilon_{\nu\lambda\alpha\beta} k_2^\alpha k_3^\beta + \hbox{\rm cyclic perms}\right)$}
\end{center}
When dotted into one of the external momenta this
reproduces (\ref{anomaly}).  The non-locality of $\Gamma$ is crucial, as otherwise the anomaly could be canceled with a local counterterm.}
\be
\label{AnomalousEA}
\Gamma[A] = \cdots \pm {q^3 \over 96 \pi^2} \int d^4x d^4y \, \partial_\mu A^\mu(x) \,
\Box^{-1}(x-y) \, \epsilon^{\alpha\beta\gamma\delta} F_{\alpha\beta} F_{\gamma\delta}(y)
\ee
In this expression $\Box^{-1}(x-y)$ should be thought of as a Green's function,
the inverse of the operator $\partial_\mu \partial^\mu$, and
$F_{\alpha\beta}$ is the field strength of $A_\mu$.
The fact that the current $j^\mu \equiv - {1 \over q} {\delta \Gamma \over \delta A_\mu}$ is not conserved,
\be
\label{divergence}
\partial_\mu j^\mu = - {1 \over q} \partial_\mu {\delta \Gamma \over \delta A_\mu} = \pm {q^2 \over 96 \pi^2}
\epsilon^{\alpha\beta\gamma\delta} F_{\alpha\beta}F_{\gamma\delta}\,,
\ee
manifests itself in the effective action as a breakdown of gauge invariance.

\section{Recovering the axial anomaly\label{sect:axial}}

Suppose we have two
spinors, one right-handed and one left-handed.  Assembling them
into a Dirac spinor $\psi$, the currents
\[
j_R^\mu = \bar{\psi} \gamma^\mu {1 \over 2} \big(1 + \gamma^5\big)\psi
\qquad\quad
j_L^\mu = \bar{\psi} \gamma^\mu {1 \over 2} \big(1 - \gamma^5\big)\psi
\]
have anomalous divergences
\be
\label{RLdiv}
\partial_\mu j^\mu_R = {q^2 \over 96 \pi^2}
\epsilon^{\alpha\beta\gamma\delta} R_{\alpha\beta} R_{\gamma\delta}
\qquad\quad
\partial_\mu j^\mu_L = - {q^2 \over 96 \pi^2}
\epsilon^{\alpha\beta\gamma\delta} L_{\alpha\beta} L_{\gamma\delta}\,.
\ee
Here $R_\mu$ and $L_\mu$ are background vector fields which couple to the
chiral components of $\psi$, and quantities with two indices are
the corresponding field strengths.  Note that we've taken the right- and
left-handed components of $\psi$ to have the same charge.  The vector and axial
currents
\[
j^\mu = j^\mu_R + j^\mu_L = \bar{\psi} \gamma^\mu \psi \qquad\quad
j^{\mu 5} = j^\mu_R - j^\mu_L = \bar{\psi} \gamma^\mu \gamma^5 \psi
\]
couple to the linear combinations
\[
V_\mu = {1 \over 2} \big(R_\mu + L_\mu\big) \qquad\quad
A_\mu = {1 \over 2} \big(R_\mu - L_\mu\big)\,.
\]
As a consequence of (\ref{RLdiv}) these currents have divergences
\beas
\partial_\mu j^\mu & = & {q^2 \over 24 \pi^2}
\epsilon^{\alpha\beta\gamma\delta} V_{\alpha\beta} A_{\gamma\delta} \\
\partial_\mu j^{\mu 5} & = & {q^2 \over 48 \pi^2}
\epsilon^{\alpha\beta\gamma\delta} \left(V_{\alpha\beta}
V_{\gamma\delta} + A_{\alpha\beta} A_{\gamma\delta}\right)\,.
\eeas
At first sight this seems no better than having a single chiral
spinor.  But consider adding the following local term to the effective
action for $V_\mu$ and $A_\mu$.
\[
\Delta \Gamma = {c q^3 \over 6 \pi^2} \int d^4x \,
\epsilon^{\alpha\beta\gamma\delta} \partial_\alpha V_\beta V_\gamma
A_\delta
\]
Here $c$ is an arbitrary constant.  This term
violates both vector and axial gauge invariance, so it contributes to
the divergences of the corresponding currents.
\beas
&& \Delta\big(\partial_\mu j^\mu\big) = - {1 \over q} \partial_\mu
{\delta(\Delta \Gamma) \over \delta V_\mu} = - {c q^2 \over 24 \pi^2}
\epsilon^{\alpha\beta\gamma\delta} V_{\alpha\beta} A_{\gamma\delta} \\
&& \Delta\big(\partial_\mu j^{\mu 5}\big) = - {1 \over q} \partial_\mu
{\delta(\Delta \Gamma) \over \delta A_\mu} = + {c q^2 \over 24 \pi^2}
\epsilon^{\alpha\beta\gamma\delta} V_{\alpha\beta} V_{\gamma\delta}
\eeas
If we add this term to the effective action and set $c = 1$,
we have a conserved vector current but an anomalous axial current.
\be
\label{AVV}
\partial_\mu j^\mu = 0 \qquad\quad
\partial_\mu j^{\mu 5} = {q^2 \over 16 \pi^2}
\epsilon^{\alpha\beta\gamma\delta} \left(V_{\alpha\beta}
V_{\gamma\delta} + {1 \over 3} A_{\alpha\beta}
A_{\gamma\delta}\right)
\ee
Given a conserved vector current we can promote $V_\mu$ to a dynamical gauge field -- usually the desired state of affairs.
If we aren't interested in making $V_\mu$ dynamical then other choices for $c$ are possible.

\section{Restoring Bose symmetry\label{sect:counterterm}}

It's no surprise that by imposing Bose symmetry we've recovered the standard expression for the anomaly, as the importance of Bose symmetry for the result
was emphasized from the very beginning \cite{Adler:1969gk,Bell:1969ts}.
But still, let's return to the momentum routing given in (\ref{TriangleAmplitude}) and see how Bose symmetry and current conservation play out.
Using identities similar to (\ref{trivial}) and the expression for the surface term (\ref{surface}) we find that
\bea
\nonumber
&& -i k_1^\mu {\cal M}_{\mu\nu\lambda} = 0 \\
&& -i k_2^\nu {\cal M}_{\mu\nu\lambda} = \mp  {q^3 \over 8 \pi^2} \epsilon_{\mu\lambda\alpha\beta} k_1^\alpha k_3^\beta \\
\nonumber
&& -i k_3^\lambda {\cal M}_{\mu\nu\lambda} = \mp {q^3 \over 8 \pi^2} \epsilon_{\mu\nu\alpha\beta} k_1^\alpha k_2^\beta
\eea
Thus with the momentum routing (\ref{TriangleAmplitude}) the current is conserved at one vertex but not at the other two, a peculiar state of affairs which shows that Bose symmetry is violated.  To capture this in an effective action we introduce three distinct vector fields $A,\,B,\,C$, with field strengths denoted by the
same letter, and take
\bea
\nonumber
\Gamma[A,B,C] & = & \pm {q^3 \over 32 \pi^2} \int d^4x d^4y \, \Big(\partial_\mu B^\mu(x) \, \Box^{-1}(x-y) \, \epsilon^{\alpha\beta\gamma\delta} A_{\alpha\beta} C_{\gamma\delta}(y) \\
\nonumber
& & \hspace{3cm} + \partial_\mu C^\mu(x) \, \Box^{-1}(x-y) \, \epsilon^{\alpha\beta\gamma\delta} A_{\alpha\beta} B_{\gamma\delta}(y)\Big) \\
\label{GammaABC}
\eea
Gauge invariance is respected for $A$ but violated for $B$ and $C$.  The breakdown of Bose symmetry is manifest.

Since Bose symmetry could be restored by symmetrizing over momentum routings, it should also be possible to restore it with a local counterterm.  Consider adding the following local
term to the effective action.
\be
\label{BoseCT}
\Delta \Gamma = \mp {q^3 \over 48 \pi^2} \int d^4x \, \epsilon^{\alpha\beta\gamma\delta} A_\alpha \left(B_{\beta\gamma} C_\delta + B_\beta C_{\gamma\delta}\right)
\ee
When added to (\ref{GammaABC}) the anomalous divergences become symmetric,
\bea
\nonumber
&&\partial_\mu j_A^\mu = - {1 \over q} \partial_\mu {\delta (\Gamma + \Delta \Gamma) \over \delta A_\mu} = \pm {q^2 \over 48 \pi^2} \epsilon^{\alpha\beta\gamma\delta} B_{\alpha\beta} C_{\gamma\delta} \\
&&\partial_\mu j_B^\mu = - {1 \over q} \partial_\mu {\delta (\Gamma + \Delta \Gamma) \over \delta B_\mu} = \pm {q^2 \over 48 \pi^2} \epsilon^{\alpha\beta\gamma\delta} A_{\alpha\beta} C_{\gamma\delta} \\
\nonumber
&&\partial_\mu j_C^\mu = - {1 \over q} \partial_\mu {\delta (\Gamma + \Delta \Gamma) \over \delta C_\mu} = \pm {q^2 \over 48 \pi^2} \epsilon^{\alpha\beta\gamma\delta} A_{\alpha\beta} B_{\gamma\delta}
\eea
and can be captured by an effective action
\be
\Gamma[A,B,C] = \pm {q^3 \over 48 \pi^2} \int d^4x d^4y \, \Big(\partial_\mu A^\mu(x) \, \Box^{-1} \, \epsilon^{\alpha\beta\gamma\delta} B_{\alpha\beta} C_{\gamma\delta}(y)  + \hbox{\rm cyclic}\Big)
\label{GammaABC2}
\ee
Then we're free to identify the three vector fields and, with a $1/3!$ for Bose symmetry, describe the anomaly with the effective action (\ref{AnomalousEA}).

The procedure above gives a consistent anomaly.  On the other hand consider adding to (\ref{GammaABC}) the counterterm\footnote{Related counterterms appear in
\cite{Paranjape:1985sk}.}
\be
\label{covariantCT}
\Delta \Gamma = \mp {q^3 \over 16 \pi^2} \int d^4x \, \epsilon^{\alpha\beta\gamma\delta} A_\alpha \left(B_{\beta\gamma} C_\delta + B_\beta C_{\gamma\delta}\right)
\ee
Thanks to the larger coefficient this counterterm squeezes all of the anomaly into one of the legs.
\be
\label{CovariantAnomaly}
\partial_\mu j_A^\mu = \pm {q^2 \over 16 \pi^2} \epsilon^{\alpha\beta\gamma\delta} B_{\alpha\beta} C_{\gamma\delta} \qquad \partial_\mu j_B^\mu = \partial_\mu j_C^\mu = 0
\ee
An effective action which captures this is
\be
\Gamma[A,B,C] = \pm {q^3 \over 16 \pi^2} \int d^4x d^4y \, \partial_\mu A^\mu(x) \, \Box^{-1} \, \epsilon^{\alpha\beta\gamma\delta} B_{\alpha\beta} C_{\gamma\delta}(y)
\label{GammaABC3}
\ee
Since $j_B$ and $j_C$ are conserved, (\ref{GammaABC3}) respects gauge invariance for $B$ and $C$, which means (\ref{CovariantAnomaly}) can be identified as the covariant anomaly for $A$.\footnote{The coefficient of (\ref{GammaABC3}) is larger than (\ref{AnomalousEA}) by a Bose factor.  Although
we can't impose full Bose symmetry on (\ref{GammaABC3}), we could set $B = C$ and divide (\ref{GammaABC3}) by $2$.  This reproduces the relative normalizations of the consistent and covariant anomalies
 in \cite{Bardeen:1984pm}.}  Evidently the covariant anomaly
can be obtained by varying an effective action, despite the Wess-Zumino consistency condition \cite{Wess:1971yu}, at the price of violating Bose symmetry.  It would be interesting to see if a similar result
holds in non-abelian theories.

\bigskip
\bigskip
\centerline{\em Happy birthday, Roman!}
\bigskip

\bigskip
\goodbreak
\centerline{\bf Acknowledgements}
\noindent
I'm grateful to V.\ Parameswaran Nair for comments on the manuscript and especially for suggesting the connection to covariant anomalies.
DK is supported by U.S.\ National Science Foundation grant PHY-1820734.


\providecommand{\href}[2]{#2}\begingroup\raggedright\endgroup

\end{document}